\documentstyle[12pt]{article}
\begin{document}

%\begin{center}
\begin{flushright}
\hfill{}
MRI-P-000901
\end{flushright}
\begin{center}
CIRCULAR COSMIC STRING LOOP IN BRANS-DICKE THEORY\\\vspace{4mm} A. Barros$^1 
$, A. A. Sen$^2$ and C. Romero$^3$\\\vspace{4mm}

$^{1}$Departamento de F\'{i}sica,\\Universidade Federal de Roraima\\69310-270%
\\Boa Vista, RR Brazil. \\\vspace{4mm} $^{2}$Mehta Research Institute\\%
Chhatnag Road, Jhusi,\\Allahabad, 211019,\\India.\\\vspace{4mm} $^{3}$%
Departamento de F\'{i}sica,\\Universidade Federal da Para\'{i}ba\\Caixa
Postal 5008, 58059-970\\Jo\~{a}o Pessoa, PB Brazil
\end{center}

\vspace{5mm} {\bf {Abstract}}

The gravitational field of a stationary circular cosmic string loop
,externally supported against collapse, is investigated in the context of
Brans-Dicke theory in the weak field approximation of the field equations.
The solution is quasi-conformally related to the corresponding solution in
Einstein's General Relativity(GR) and goes over to the corresponding
solution in GR when the Brans-Dicke parameter $\omega $ becomes infinitely
large. \\

\vspace{5mm} PACS NO: 04.20Jb, 98.80.Cq \\\vspace{2mm} % \newpage

Phase transition in the early universe might have produced some topological
defects\cite{kib}. Amongst these defects, cosmic strings, have attracted a
lot of interest for various reasons\cite{vil}. For instance, they are
capable of producing observational effects such as double images of quasars
and are considered to be possible candidates as seeds for galaxies
formation.

In some gauge models strings do not have any end, and thus are either of
infinite extent form or closed circular loops. One of the most notable
features of the gravitational field of a straight infinite cosmic string is
the presence of an ''angular deficit'', in an otherwise Minkowskian
spacetime, having a magnitude related to the linear energy density $\mu $ of
the string by the equation $\delta \Phi =8\pi G\mu $. In fact, this angular
deficit plays a key role for the production of double images of quasars\cite
{loc}.

The deficit angle model is widely believed to be a good approximation for
describing a spacetime exterior to the string core. Frolov, Israel and Unruh
(FIU)\cite{fiu} used this approximation for studying a closed circular
cosmic string loop at a moment of time symmetry. With the help of the
initial value formulation\cite{adm}, they produced a family of momentarily
stationary circular cosmic strings, which are regarded as thin loops either
at the time of formation or at the turning point of expansion and collapse.
An important assumption in this work is that all points on the circular
string be conical singularities with angular deficit equal to that of a
infinite straight string of equal linear energy density. Hughes, McManus,
and Vandyck(HMV)\cite{hmv} investigated further the problem of angular
deficit of a circular string. They considered a weak field stationary
solutions of Einstein's field equations for a thin circular string and also
established that radial stress should be introduced to support the string
loop against possible gravitational collapse. They determined the form of
the radial stress from the stress energy conservation relations. The main
result of their study is that in weak field approximation a circular string
produces, locally, along the string the same angular deficit as does a
straight infinitely long string with the same linear energy density. In
another work, McManus and Vandyck\cite{mv} considered a string loop with a
rotation of the loop which provides the necessary centrifugal reaction
partially or fully in order to avoid the possible gravitational collapse.

It turns out that at sufficient high energy scales it seems likely that
gravity is not given by the Einstein's action, but becomes modified by the
superstring terms. In the low energy limit of this string theory one
recovers Einstein's gravity along with a scalar dilaton field which is
non-minimally coupled to gravity~\cite{gsw}. On the other hand, scalar
tensor theories, such as Brans-Dicke theory(BD)~\cite{BD}, have been
considerably revived in recent years. It was shown by La and Steinhardt~\cite
{LS} that because of the interaction of the BD scalar field with the Higgs
type sector, the exponential inflation in Guth's model ~\cite{Guth} could be
slowed down to power law one and the graceful exit in the inflation is thus
completed via bubble nucleation. Although dilaton gravity and BD theory
arise from entirely different motivations, it can be shown that the former
is a special case of the latter, at least formally~\cite{AAS}. Another
motivation for studying gravitational properties of defects in BD theory is
that the only defects we can hope to observe now are those formed after or
near the end of inflation, and the formation of such superheavy defects is
relatively easy to arrange in a Brans-Dicke type theory~\cite{CKL}.

In the present work we have studied the gravitational field of a stationary
circular cosmic string loop in Brans-Dicke theory in the weak field
approximation of the field equations.

The field equations in the BD theory are written in the form 
$$
G_{\mu \nu }=8\pi {\frac{T_{\mu \nu }}{{\phi }}}+{\frac \omega {{\phi ^2}}}%
(\phi _{,\mu }\phi _{,\nu }-{\frac 1{{2}}}g_{\mu \nu }\phi _{,\alpha }\phi
^{,\alpha })+{\frac 1{{\phi }}}(\phi _{,\mu ;\nu }-g_{\mu \nu }\Box \phi ),%
\eqno{(1.1)} 
$$
$$
\Box \phi ={\frac{8\pi T}{{2\omega +3}}},\eqno{(1.2)} 
$$
where $\phi $ is the scalar field, $\omega $ is the BD parameter and $T$
denotes the trace of the energy-momentum tensor $T_\nu ^\mu $~\cite{BD}. In
the weak field approximation of BD theory one can assume $g_{\mu \nu }=\eta
_{\mu \nu }+h_{\mu \nu }$ where $|h_{\mu \nu }|<<1$ and $\phi (r)=\phi
_0+\epsilon (r)$ with $|\epsilon /\phi _0|<<1$ where $1/\phi _0=G_0={\frac{%
(2\omega _{+}3)}{{(2\omega +4)}}}G$ in order to have a Newtonian limit for
the BD theory\cite{BD}.

It has been shown recently by Barros and Romero~\cite{BR2}, that in the weak
field approximation the solutions of the BD equations are related to the
solutions of linearized equations in GR with the same $T_\nu ^\mu $ in the
following way: if $g_{\mu \nu }^{gr}(G,x)$ is a known solution of the
Einstein's equations in the weak field approximation for a given $T_\nu ^\mu 
$, then the BD solution corresponding to the same $T_\nu ^\mu $, in the weak
field approximation, is given by 
$$
g_{\mu \nu }^{bd}(x)=[1-G_0\epsilon (x)]g_{\mu \nu }^{gr}(G_0,x)\eqno{(1.3)} 
$$
where $\epsilon (x)$ must satisfy 
$$
\Box \epsilon (x)={\frac{8\pi T}{{(2\omega +3)}}},\eqno{(1.4)} 
$$
and $G$ is replaced by $G_0$ defined previously. Hence, to get the spacetime
for circular string loop one has to solve the equation (1.4) with
appropriate $T_\nu ^\mu $ for the circular cosmic string loop.

We have taken the form of the energy momentum tensor for the string loop as
proposed by Hughes et al \cite{hmv}: 
$$
T_t^t=-\mu \delta (r-a)\delta (z)\eqno{(1.5a)} 
$$
$$
T_\phi ^\phi =k\delta (r-a)\delta (z)\eqno{(1.5b)} 
$$
$$
T_r^r={\frac k{{r}}}\Theta (r-a)\delta (z)\eqno{(1.5c)} 
$$
in which $\Theta $ denotes the Heaviside step function. Here we are
considering an infinitely thin loop of string with radius $a$, lying in the $%
x$-$y$ plane and centered at the origin. For a circular string $T_\phi ^\phi 
$ plays the same role as that of the longitudinal stress $T_z^z$ for a
straight string and $T_r^r$ is the external radial stress required for
supporting the loop against collapse and hence it is not localized on the
string.

With (1.5), equation (1.4) becomes

$$
\nabla ^2\epsilon =\frac{8\pi }{2\omega +3}[(k-\mu )\delta \left( r-a\right)
\delta (z)+\frac kr\Theta (r-a)\delta (z)]\eqno{(1.6)}
$$
Writing $\epsilon =\epsilon _1+\epsilon _2$, we can separate equation (1.6)
into 
$$
\nabla ^2\epsilon _1=\frac{8\pi }{2\omega +3}[(k-\mu )\delta (r-a)\delta (z)]%
\eqno{(1.7a)}
$$
$$
\nabla ^2\epsilon _2=\frac{8\pi }{2\omega +3}[\frac kr\Theta (r-a)\delta (z)]%
\eqno{(1.7b)}
$$
The solutions of the equations above can be found easily by using toroidal
coordinates $(\phi ,\sigma ,\psi )$ which are related to cylindrical
coordinates $(\phi ,r,z)$ by \cite{fiu} 
$$
z=aN^{-2}\sin \psi \eqno{(1.8a)}
$$
$$
r=aN^{-2}\sinh \sigma \hspace{2mm}(0\leq \sigma \leq \infty ,\left| \psi \right|
\leq \pi )\eqno{(1.8b)}
$$
where $N^2\equiv N^2(\sigma ,\psi )\equiv \cosh \sigma -\cos \psi $. The
surfaces $\sigma =\sigma _0\equiv $ constant are tori whose generating
circles have radii $a$csch $ \sigma _0$ and $a\coth \sigma _0$. The ring $r=a$%
, $z=0$ is now given by $\sigma =\infty $.

In toroidal coordinates the solution of equation (1.7a) now becomes\cite{bat}
$$
\epsilon _1=-2^{\frac 52}\frac{(k-\mu )}{(2\omega +3)}N(\sigma ,\psi )\frac{%
F(\tanh (\sigma /2))}{\cosh (\sigma /2)}\eqno{(1.9a)} 
$$
where $F$ denotes complete elliptic integral of the first kind and the
solution for $\epsilon _2$ becomes\cite{hmv} 
$$
\epsilon _2=-\frac{4k}{(2\omega +3)}N(\sigma ,\psi )\sum\limits_{n=0}^\infty
[H_n(\sigma )P_{n-1/2}(\cosh \sigma )+G_n(\sigma )Q_{n-1/2}(\cosh \sigma
)]\varepsilon _n\cos (n\psi )\eqno{(1.9b)} 
$$
where 
$$
H_n(\sigma )\equiv \int\nolimits_0^\sigma dxN^{-1}(x,0)Q_{n-1/2}(\cosh x)%
\eqno{(2.0a)} 
$$
$$
G_n(\sigma )\equiv \int\nolimits_0^\sigma dxN^{-1}(x,0)P_{n-1/2}(\cosh x)%
\eqno{(2.0b)} 
$$
and $P_{n-1/2}^m$ and $Q_{n-1/2}^m$ are toroidal Legendre functions, $%
\varepsilon _n\equiv 2-\delta _n^0$, and also we have used the notation $%
P_{n-1/2}\equiv P_{n-1/2}^0$ and $Q_{n-1/2}\equiv Q_{n-1/2}^0$. Thus, from
equation (1.3) the spacetime for a stationary circular cosmic string loop in
BD theory in the weak field approximation is given by 
$$
ds^2=[1-G_0(\epsilon _1+\epsilon _2)]ds_{HMV}^2(G_0)\eqno{(2.1)} 
$$
where $\epsilon _{1}$and $\epsilon _2$ are given by equations (1.9a)
and (1.9b), and $ds_{HMV}^2(G_0)$ is the metric obtained by Hughes et al\cite
{hmv} for circular string loop in GR with $G$ replaced by $G_0$.

To investigate the presence of conical singularities on the ring $\sigma
=\infty $, we proceed as follows: first of all, for the scalar field
solution in equation (1.9) we may add an arbitrary constant $A$ such that 
$$
\epsilon =\epsilon _1+\epsilon _2+A
\eqno{(2.2)} 
$$
For $\sigma \rightarrow \infty $ one can find the asymptotic behaviour of $%
\epsilon _1$ and $\epsilon _2$ from equations (1.9) which is given by\cite
{hmv} 
$$
\epsilon _1\rightarrow -{\frac{4(k-\mu )\sigma }{{(2\omega +3)}}}%
\eqno{(2.3a)} 
$$
$$
\epsilon _2\rightarrow 0
\eqno{(2.3b)} 
$$
Hence, we have 
$$
1-G_0\epsilon =1+{\frac{4G_0(k-\mu )\sigma }{(2\omega +3)}}-G_0A\eqno{(2.4)} 
$$
Then, for the surface $t=$ constant and $\phi =$ constant and using the
toroidal coordinates one can obtain for the metric (2.1)\cite{hmv} 
$$
ds^2=[1+{\frac{4G_0(k-\mu )\sigma }{(2\omega +3)}}-G_0A][4a^2e^{-2\sigma
}e^{4G_0(\mu -k)\sigma +2b}(d\sigma ^2+d\psi ^2)]\eqno{(2.5)} 
$$
where $b$ is the dimensionless combination of all additive constants
appearing in the solutions . Now, let us make the following coordinate
transformation: 
$$
\psi =\theta -\pi \eqno{(2.6a)} 
$$
$$
2ae^{-(\sigma +2G_0(k-\mu )\sigma -b)}=r\eqno{(2.6b)} 
$$
Given that $b$ is a first order term in this approximation \cite{hmv} we can
write 
$$
1+{\frac{4G_0(k-\mu )\sigma }{(2\omega +3)}}-G_0A=1+{\frac{4G_0(\mu -k)}{%
(2\omega +3)}}\ln (r/2a)-G_0A\eqno{(2.7)} 
$$
By choosing $A={\frac{4(\mu -k)}{(2\omega +3)}}\ln (r_0/2a)$, where $r_0$ is
a constant, we get 
$$
ds^2=[1+{\frac{4G_0(\mu -k)}{(2\omega +3)}}\ln (r/r_0)][{\frac{dr^2}{%
(1-4G_0(\mu -k))}}+r^2d\theta ^2]\eqno{(2.8)} 
$$
Defining $dr^{^{\prime }}={\frac{dr}{\sqrt{1-4G_0(\mu -k)}}}$, one can write
the above equation as 
$$
ds^2=[1+{\frac{4G_0(\mu -k)}{(2\omega +3)}}\ln (r^{^{\prime
}}/r_0)][dr^{^{\prime }2}+r^{^{\prime }2}(1-4G_0(\mu -k))d\theta ^2]%
\eqno{(2.9)} 
$$
Finally, if we put $k=-\mu $ for ``string'' matter we are led to 
$$
ds^2=[1+{\frac{8G_0\mu }{(2\omega +3)}}\ln (r^{^{\prime
}}/r_0)][dr^{^{\prime }2}+(1-8G_0\mu )r^{^{\prime }2}d\theta ^2]\eqno{(2.10)}
$$
This is exactly the same metric for a straight vacuum string for section $t=$
constant and $z=$ constant which was earlier obtained by Barros and Romero%
\cite{BR3} in BD theory in the weak field approximation. At this point, it
is worthwhile mentioning that this spacetime has an angular deficit $\Delta
\Phi $ given by 
\[
\Delta \Phi =8\pi G_0\mu \left[ 1-\frac 1{2\omega +3}\ln \left( \frac{%
r^{^{\prime }}}{r_0}\right) \right] 
\]
as may be seen directly by integrating (2.10) around a circle of radius $%
r^{\prime }$.

Finally, it is well known that in the weak field approximation when $\omega
\rightarrow \infty $ the BD solution goes over to the corresponding solution
in Einstein's GR, although this is not always true in the case of exact
solutions\cite{BRS}. In our case one can check that for $\omega \rightarrow
\infty $, both $\epsilon _1$and $\epsilon _2$ become zero and $%
G_0\rightarrow G$, so one can recover the corresponding GR solution for a
circular cosmic string obtained by Hughes et al\cite{hmv}.

In conclusion, we have obtained the spacetime for circular cosmic string
loop in BD theory in the weak field approximation. In doing so, we have
followed the method prescribed by Barros and Romero\cite{BR2}. When the loop
is made of ``string'' matter we find out that at points near the string ($%
\sigma \rightarrow \infty $) we recover the result previously obtained by
Barros and Romero\cite{BR3} for a straight static local string in BD theory.
The solution goes over to the corresponding solution in GR in the limit $%
\omega \rightarrow \infty $.\\


\newpage
\begin{thebibliography}{99}
\bibitem{kib}  T. W. Kibble, J. Phys. A {\bf 9}, 1387 (1976).

\bibitem{vil}  A. Vilenkin, Phys. Rep. {\bf 121}, 263 (1985).

\bibitem{loc}  A. Vilenkin, Phys. Rev. D {\bf 24}, 2082 (1981); J. R. Gott,
Astrophys. J. {\bf 288}, 422 (1985); T. Vachaspati and A. Vilenkin, Phys.
Rev. D {\bf 31}, 3052 (1985); R. Brandenberger, A. Albrecht and N. Turok,
Nucl. Phys. B {\bf 277}, 605 (1986).

\bibitem{fiu}  V. P. Frolov, W. Israel and W. G. Unruh, Phys. Rev. D {\bf 39}%
, 1084 (1989).

\bibitem{adm}  R. Arnowitt, S. Deser and C. W. Misner, in: {\bf Gravitation:
An Introduction to Current Research}, edited by L. Witten (Wiley, N. Y.,
1962).

\bibitem{hmv}  S. J. Hughes, D. J. McManus and M. A. Vandyck, Phys. Rev. D 
{\bf 47}, 468 (1993).

\bibitem{mv}  D. J. McManus and M. A. Vandyck, Phys. Rev. D {\bf 47}, 1491
(1993).

\bibitem{gsw}  M. B. Green, J. H. Schwartz and E. Witten, {\bf Superstring
Theory }(Cambridge University Press, Cambridge, England, 1987).

\bibitem{BD}  C. Brans and R. H. Dicke, Phys. Rev. {\bf 124}, 925 (1961).

\bibitem{LS}  D. La and P. J. Steinhardt, Phys. Rev. Lett. {\bf 62}, 376
(1989).

\bibitem{Guth}  A. H. Guth, Phys. Rev. D {\bf 23}, 347 (1981).

\bibitem{AAS}  A. A. Sen, Phys. Rev. D {\bf 60}, 067501 (1999).

\bibitem{CKL}  E. J. Copeland, E. W. Kolb and A. Liddle, Phys. Rev. D {\bf 42%
}, 2911 (1990).

\bibitem{BR2}  A. Barros and C. Romero, Phys. Lett. A {\bf 245}, 31 (1998).

\bibitem{bat}  H. Bateman, {\bf Partial Differential Equations of
Mathematical Physics }(Cambridge University Press, Cambridge, England, 1952).

\bibitem{BR3}  A. Barros and C. Romero, J. Math. Phys. {\bf 36}, 5800 (1995).

\bibitem{BRS}  C. Romero and A. Barros, Phys. Lett. A {\bf 173}, 243 (1993);
N. Banerjee and S. Sen, Phys. Rev. D {\bf 56}, 1334 (1997).
\end{thebibliography}
\end{document}